\documentclass[prl,twocolumn]{revtex4}
\usepackage{epsfig}
\usepackage{bm}

\begin{document}

\title{Hamiltonian dynamics and distributed chaos in DNA}

\author{A. Bershadskii}

\affiliation{
ICAR, P.O. Box 31155, Jerusalem 91000, Israel
}

\begin{abstract}
It is shown that distributed chaos, generated by Hamiltonian DNA dynamics with spontaneously broken {\it time} translational symmetry, imprints itself on the DNA sequence of Arabidopsis thaliana (a model plant for genetic sequencing and mapping) and of the NRXN1 and BRCA2 human genes (as an example). The base-stacking interactions in the DNA duplex and degenerate codon groups have been discussed in this context.

\end{abstract}

\maketitle

\section{Introduction}
 
Order of the four nucleotide bases: adenine (A), cytosine (C), guanine (G), and thymine (T) on a DNA strand (so-called DNA sequence) is used by a DNA molecule to encode the biological information. The nucleotides A (adenine) at one strand of the double-stranded DNA duplex always pairs with nucleotides T (thymine) at another strand, and nucleotides G (guanine) always pairs with nucleotides C (cytosine). The DNA's helix axis inside a cell's nucleus is strongly curved. Human genome, for instance, has the length approximately 1 meter, while the nucleus of a human somatic cell has diameter approximately $5 \times 10^{-6}$m. Due to the very confined physical conditions for the DNA molecules inside cells' nucleus their information encoding should be adjusted (at least statistically) to physical dynamics of the nucleotide sequences on their strands. On the other hand, the non-linear waves running along the DNA strands can be described using Hamiltonian approach \cite{e},\cite{pb} (see for a comprehensive review  Ref. \cite{y}). The Hamiltonian non-linearity is a basis for chaotic dynamics \cite{suz}. Therefore, one can expect an imprint of the chaotic dynamics on the DNA
sequence. In order to study this phenomenon it was suggested in Ref. \cite{b1} to map the DNA sequence to telegraph signals. But before using this mapping let us look at the chaotic properties of the Hamiltonian dynamic systems, which we expect to observe in this case. 

\section{Distributed chaos}

 Chaotic behaviour in dynamical systems is often characterized by exponential power spectra \cite{oh}-\cite{fm}
$$
E(f) \propto \exp -(f/f_c)      \eqno{(1)}
$$
where $f_c = const$ is some characteristic frequency. It is shown in the Ref. \cite{b1} that in the DNA sequences positions of energy minima of stacking interactions are controlled by a chaotic order characterized by such spectrum.

   A more complex (distributed) chaos has the spectra which can be approximated by a weighted superposition of the exponentials Eq. (1):
$$
E(f) \propto \int_0^{\infty} P(f_c)~ \exp -(f/f_c)~ df_c   \eqno{(2)}
$$
where $f_c$ has a probability distribution $P(f_c )$.

  The energy conservation law is related to the time translational symmetry by the Noether's theorem \cite{ll}. For the systems with spontaneously broken time translational symmetry the action $I$ is an adiabatic invariant \cite{suz}. For such systems the characteristic frequency $f_c$ can be related to characteristic velocity $v_c$ by the dimensional considerations:
$$    
v_c \propto I^{1/2} f_c^{1/2}  \eqno{(3)}
$$
Hence normal (Gaussian) distribution of the characteristic velocity results in the chi-squared distribution ($\chi^{2}$) of $f_c$
$$
P(f_c) \propto f_c^{-1/2} \exp-(f_c/4f_0),  \eqno{(4)}
$$
$f_0$ is a constant. 

   Substitution of the $\chi^{2}$ distribution Eq. (4) into integral Eq. (2) results in
$$
E(f) \propto \exp-(f/f_0)^{1/2}  \eqno{(5)}
$$

\section{DNA sequence mapping}

Since the Hamiltonian dynamics of the the DNA sequence should imprint itself on the DNA information encoding (see Introduction), let us map the 'A-C-G-T' language of the DNA sequence into a set of telegraph signals in order to see this imprinting. Let us define binary functions $B_A(n)$,  $B_C(n)$, $B_G(n)$, $B_T(n)$ of natural numbers $n = 1, 2, 3,...$ ($n$ enumerates consequent places in the DNA sequence). The telegraph signal $B_A(n)$, for instance, corresponds to the letter A in the DNA sequence. Such binary function takes two values +1 or -1 and changes its sign passing each letter A along the DNA sequence (the functions $B_C(n)$, $B_G(n)$, and $B_T(n)$ are analogously defined).
   We have mapped into the telegraph signals the genome sequence of the flowering plant Arabidopsis thaliana, which is a model plant for genome analysis \cite{mei}. Its genome is one of the smallest plant genomes that makes Arabidopsis thaliana useful for genetic sequencing
and mapping. The most up-to-date version of the Arabidopsis thaliana genome is maintained by The Arabidopsis Information Resource (TAIR) \cite{plant}.
\begin{figure} \vspace{-1.5cm}\centering
\epsfig{width=.45\textwidth,file=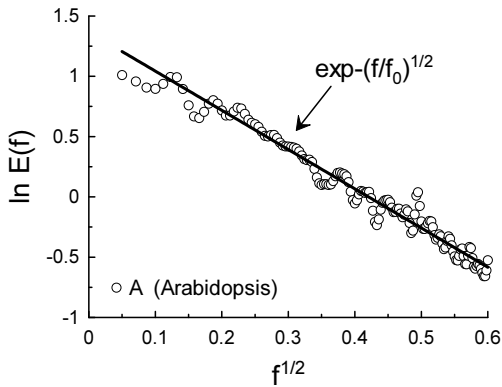} \vspace{-4.5cm}
\caption{Power spectrum of the $B_A(n)$ signal. The solid straight line indicates the stretched exponential decay Eq. (5).}
\end{figure}
\begin{figure} \vspace{-0.5cm}\centering
\epsfig{width=.45\textwidth,file=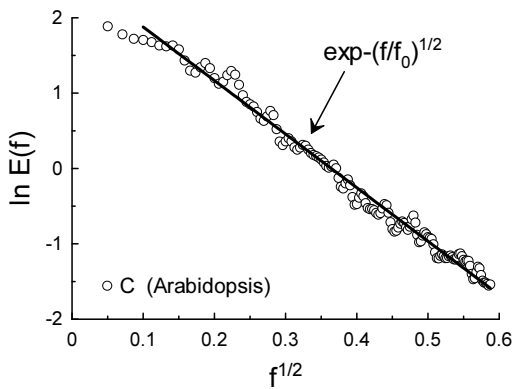} \vspace{-4cm}
\caption{As in Fig. 1 but for the $B_C(n)$ signal.}
\end{figure}
    
    Figure 1 shows power frequency spectrum $E(f)$ for the telegraph signal $B_A(n)$ computed using a DNA sequence of the Arabidopsis thaliana. The computation was made with the maximum
entropy method (this method provides optimal spectral
resolution for the spectra caused by chaotic phenomena \cite{oh}). The semi-logarithmic scales are used in order to compare the data with the Eq. (5). The straight line is drawn in the Fig. 1 to indicate correspondence of the data to the Eq. (5) ($T_0 =1/f_0 \simeq 11$). Analogous spectrum computed for $B_C(n)$ has been shown in Fig. 2 ($T_0 =1/f_0 \simeq 49$). Figure 3 shows power spectrum of the $B_A(n)$ for human gene NRXN1 (for the genome sequence see Ref. \cite{nih}). This gene molecular location on chromosome 2 corresponds to base pairs 50,145,642 to 51,259,673 (one of the largest human genes related, in particular, to nicotine dependence and to susceptibility to autism). The straight line in Fig. 3 indicates correspondence to the Eq. (5) ($T_c=1/f_c \simeq 13$).

\section{Stacking interaction in DNA duplex}

In the double-stranded DNA duplex the bases pairs  A - T, as well as G
- C, form effective hydrogen bonds. Therefore, A on one strand always pairs with T on the opposite strand, and G always pairs with C. On the other hand, the stacking interactions of the bases pairs come from hydrophobic interactions and from the overlap of the $\pi$ electrons of the bases, and the combinations TA and AT indicate positions of stacking energy minima at DNA sequence \cite{b1}. 
Since the base-stacking interactions contribute significantly into the DNA duplex stability, it is important to look also at the distribution of the  positions of stacking energy minima at DNA sequence. 

  For the energy minima mapping the multiple AT/TA combinations in the DNA sequence should be considered as a single letter \cite{b1}, say Z (e.g. the multiple combination TAATTATA $\rightarrow$ Z). Let us define $\tilde{B}_{AT/TA}(n) = B_Z (n)$, where index AT/TA means any multiple AT/TA combination (cluster) at the sequence.  
\begin{figure} \vspace{-1.5cm}\centering
\epsfig{width=.45\textwidth,file=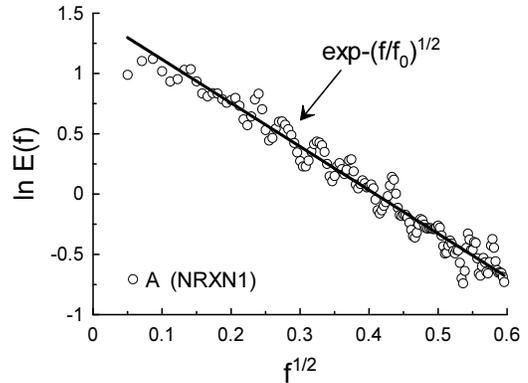} \vspace{-4.3cm}
\caption{As in Fig. 1 but for the NRXN1 human gene.}
\end{figure}
\begin{figure} \vspace{-0.3cm}\centering
\epsfig{width=.45\textwidth,file=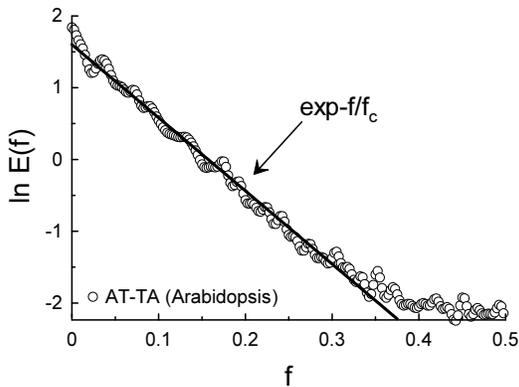} \vspace{-5.3cm}
\caption{Power spectrum of the $\tilde{B}_{AT/TA}(n)$ signal for Arabidopsis. The solid straight line indicates the exponential decay Eq. (1).}
\end{figure}

\begin{figure} \vspace{-0.5cm}\centering
\epsfig{width=.45\textwidth,file=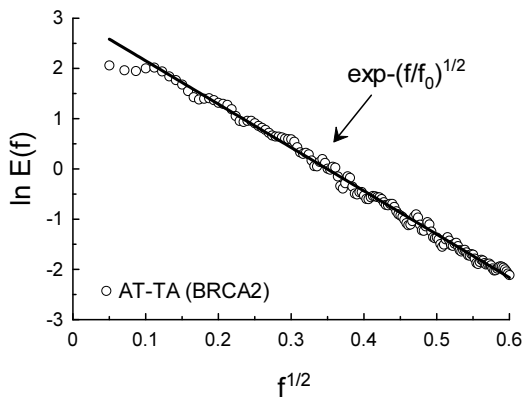} \vspace{-4.05cm}
\caption{Power spectrum of the $\tilde{B}_{AT/TA}(n)$ signal for BRCA2. The solid straight line indicates the stretched exponential decay Eq. (5).}
\end{figure}

It is shown in the Ref. \cite{b1} that power spectrum of the $\tilde{B}_{AT/TA}(n)$ for the Arabidopsis plant is exponential, that corresponds to the deterministic chaos - Eq. (1) (Fig. 4, $T_c=1/f_c \simeq 10$). The power spectrum of the $\tilde{B}_{AT/TA}(n)$ for human gene NRXN1 is rather similar to that of the Arabidopsis plant shown in the Fig. 4 (with $T_c \simeq 13$). On the other hand, figure 5 shows power spectrum of the $\tilde{B}_{AT/TA}(n)$ for human gene BRCA2 (for the genome sequence see Ref. \cite{nih}). This gene molecular location on chromosome 13 corresponds to base pairs 32,889,616 to 32,973,808. The BRCA2 gene is a caretaker responsible for genome stability. It produces tumor suppressor proteins and prevents an uncontrolled way of the cells dividing and growing (i.e. mutations in the BRCA2 gene are strongly related to a risk of cancer). The straight line in Fig. 5 indicates correspondence to the Eq. (5) (the distributed chaos, $T_0=1/f_0 \simeq 75$).  Comparing the Fig. 4 ($T_c \simeq 10$) and Fig. 5 ($T_0 \simeq 75$) one can conclude that the stacking interactions are much more long-range for the distributed chaos.

\section{Degenerate codon groups}

In the process of making proteins the alphabet of the DNA (A,C,G,T) is translated into alphabet of mRNA (triplet codons) and then, according to the genetic code, into alphabet of proteins. Not all nucleotide bases consisting the protein-coding gene regions of a DNA sequence then appear at the corresponding mRNA sequences. Some of them, belonging to the pieces of the DNA sequence called introns, are spliced out at the transcription process. The remaining pieces of the DNA sequence (called exons) are ligated into the mRNA. Therefore storage of the  
 nucleotide triplets in the DNA sequences (potential mRNA codons) is much more abundant than that of the mRNA. Order of the remaining triplets is the same as in the DNA sequence (with substituting thymine-T by uracil-U). Degeneration of the genetic code is another type of redundancy. Namely, most amino acids are specified in this code by more than one codon. The most degenerate (6-fold) codon groups correspond to Leucine (Leu), Serine (Ser) and Arginine (Arg). For instance, in the DNA terms TTA,TTG,CTT,CTC,CTA, and CTG form the Leucine (Leu) degenerate codon group. Due to this redundancy, in particular, an error in the third position of the triplet does not affect the protein (a silent mutation). There is a physico-chemical affinity between the cognate nucleotide triplets and the produced amino acids (see, for instance, Ref. \cite{kn}). This affinity allows to consider the nucleotide triplets from a degenerate codon group as a physico-chemical family. If this 'family' property survives the T$\longleftrightarrow$U replacement at the transcription process, then the abundant storage of the corresponding DNA nucleotide triplets (potential codons) also forms a physico-chemical family and, consequently, should have an imprint of the distributed chaos (see Introduction). Figure 6 shows power spectrum of the $\tilde{B}(n)$ signal computed for the \{TTA,TTG,CTT,CTC,CTA, and CTG\} family of the DNA nucleotide triplets (BRCA2 gene) corresponding to Leucine (Leu) (see above). The straight line in Fig. 6 indicates correspondence to the Eq. (5) (the distributed chaos, $T_0=1/f_0 \simeq 92$).
\begin{figure} \vspace{-1.4cm}\centering
\epsfig{width=.45\textwidth,file=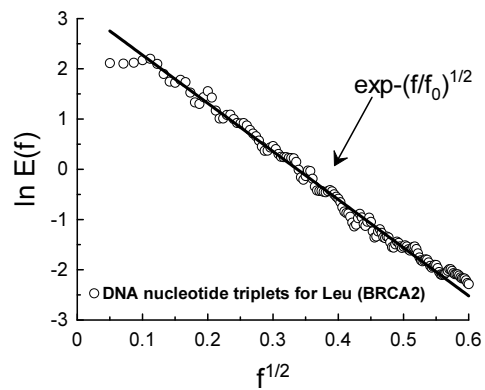} \vspace{-4.17cm}
\caption{Power spectrum of the $\tilde{B}(n)$ signal computed for the \{TTA,TTG,CTT,CTC,CTA, and CTG\} family of the DNA nucleotide triplets corresponding to Leucine (Leu). }
\end{figure}

\begin{figure} \vspace{-0.4cm}\centering
\epsfig{width=.45\textwidth,file=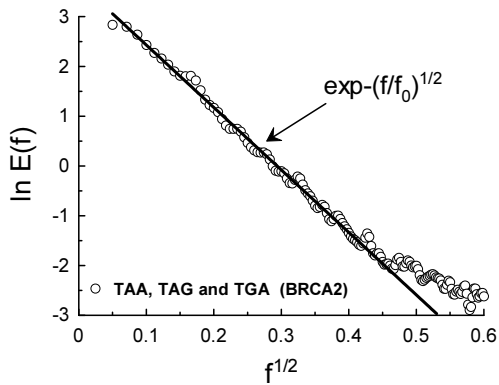} \vspace{-4.15cm}
\caption{As in Fig. 6 but for the \{TAA,TAG and TGA\}.}
\end{figure}
  Another type of the degenerate codon group is the stop codon group: in the DNA terms TAA, TAG and TGA. These codons also have a common function in the genetic code - they give a signal to stop the protein synthesis. If the stop nucleotide triplets also form a physico-chemical family in the above mentioned sense, then the power spectrum of the $\tilde{B}(n)$ signal computed for this DNA family of the nucleotide triplets \{TAA,TAG and TGA\} should have an imprint of the distributed chaos (Fig. 7 for the BRCA2 gene). The straight line in Fig. 7 indicates correspondence to the Eq. (5) (the distributed chaos, $T_0=1/f_0 \simeq 158$).

\section{Acknowledgement}

I acknowledge using the data provided by the sites: www.plantgdb.org and www.ncbi.nlm.nih.gov.

\end{document}